\documentclass[aps,twocolumn,
showkeys,showpacs,nofootinbib,byrevtex]{revtex4-1}
\usepackage{amsmath,amsfonts,amssymb,amscd,amsxtra,amsthm}
\usepackage{graphicx}
\usepackage{bm}
\usepackage{ulem}
\usepackage{epsfig}

\begin{document}

\title{A new combined analysis of elastic and charge exchange $K N(\overline{K}N)$ scatterings
in the Regge realm}

\author{Byung-Geel Yu}%
\email{bgyu@kau.ac.kr}%
\affiliation{Research Institute of Basic Science, Korea Aerospace
University, Koyang, 10540, Korea}

\author{Kook-Jin Kong}%
\email{kong@kau.ac.kr}%

\date{\today}

\begin{abstract}
The Regge features of elastic $K^\pm N\to K^\pm N$, and charge
exchange $K^-p\to\overline{K}^0n$ and $K^+n\to K^0p$ reactions are
described by using a combined analysis of all these channels at
high momenta. Based on the meson exchanges in the $t$-channel with
their decays to $K\overline{K}$ pair allowed, the exchanges of
$\rho(775)+\omega(782)$ are excluded in contrast to existing model 
calculations and the present work follows a new scheme for the meson exchanges
$a_0(980)+\phi(1020)+f_2(1275)+a_2(1320)$ reggeized for the
forward elastic scattering amplitude together with Pomeron. The
charge exchange reactions are described by $a_0(980)+a_2(1320)$
exchanges in the $t$-channel. Dominance of the isoscalar $f_2$ and
Pomeron exchanges beyond $P_{\rm Lab}\approx3$ GeV/$c$ is shown in
the elastic process. Both the isovector $a_0$ and $a_2$ exchanges
in charge exchange reactions play the respective roles in the low
and high momentum region. Differential and total cross sections
are presented to compare with existing data. Discussion is given
to the polarization of $K^-p\to K^-p$ at $P_{Lab}=10$ GeV/$c$ and
$K^-p\to \overline{K}^0n$ at $P_{Lab}=$ 8 GeV/$c$.
\end{abstract}
\maketitle

\section{introduction}

Meson-baryon scatterings are the most fundamental reactions to
understand strong interaction purely on hadronic bases. Among
these reactions, $KN$ and $\overline{K}N$ scatterings are
interesting, because they could offer a testing ground for the
formation of exotic baryons in the $K^+N$ reaction
\cite{giacomelli,arndt,azimov,hyslop} as well as the kaonic bound
state through the $K^-N$ channel coupling \cite{aoki,ikeda}. Above
the region, $P_{\rm Lab}\geq 3$ GeV/$c$, of course, the
$t$-channel meson exchange becomes dominant and in the elastic
process, in particular, the Pomeron exchange is expected to yield
a nonresonant diffraction toward high momenta up to hundreds of
GeV/$c$ \cite{donnachie}.

To date, however, in comparison to an effort to understand the
reaction mechanism at low momentum \cite{ikeda,oset}, the
description of such an high momentum aspect of $KN$
$(\overline{K}N)$ scattering is less challenged, since after the
early stage of the Regge theory with the residue fitted to
empirical data \cite{irving}. It has been a convention that the
exchanges of $\rho$ and $\omega$ mesons are included to give
contributions based on the combined fitting procedure to $\pi N$
and $KN$ data with their coupling strengths from the $SU(3)$
symmetry \cite{phillips,rarita,butt,nys,bszou}, On the other hand,
from their decay modes to $K\overline{K}$ which are forbidden
kinematically,  these lighter vector mesons are highly
off-shell and, thus, their coupling constants should also be
highly model-dependent, if participating in the reaction. This is
true for lighter meson exchanges in the $NN$ scattering. In
practice, the former reaction proceeds via the exchanged meson
decaying to $K\overline{K}$ in the $t$-channel, whereas the latter
case undergoes exchange of strong force between two nucleon
lines by the exchange of virtual mesons with hadron form factors.
In these respect, the meson-baryon scattering could be treated
differently from the baryon-baryon scattering, when the virtual
mesons should decay to the on shell $K\overline{K}$ pair,
as are the cases of $\rho$ and $\omega$.

Further, we recall that in the total (reaction) cross
sections of $K^\pm p\to K^\pm p$ \cite{donnachie} the roles of
$\rho$ and $\omega$ exchanges are unique to distinguish between
$K^+p$ and $K^-p$ cross sections at high momenta, though far away
from their on-shell propagations. However, such a conspicuous
difference is not observed in experimental data between elastic
$K^\pm p$ scatterings. Rather, they are coincident to each other,
in contrast to
the case of total cross section. Thus, we are led to consider
a single Pomeron exchange without $\rho+\omega$ exchanges for
the elastic $K^\pm p$ phenomena at high momenta.
This issue will be pointed out in the
Appendix with a demonstration for the inconsistency of the
$\rho+\omega$ exchanges with the elastic $K^\pm p$ cross
sections.

In hadron models which are based on the on-shell Born amplitudes
such as the standard pole model, or its Reggeized version,
it is hard to employ these lighter vector mesons without either a
large uncertainty in their coupling strengths or
a model dependence due to the cutoff form factors.
Therefore, with a question about how the theory
without $\rho+\omega$ exchanges could work on the elastic
$KN\ (\overline{K}N)$ scattering, it is worth investigating the
reaction based on a new scheme of the $t$-channel
meson exchange with those that are decaying to $K\overline{K}$
as reported in the
Particle Data Group (PDG).

In our previous work \cite{yu-kong-pin} we investigated forward
scattering of $\pi N$ elastic and charge exchange processes in the
Regge model where the relativistic Born terms for the $t$-channel
mesons were reggeized  with the interaction Lagrangians and
coupling constants sharing with those widely accepted in other
hadron reactions. The diffractive features of elastic reactions up
to hundreds of ${\rm GeV}/c$ pion momentum were well described by
the Pomeron exchange that we constructed from the quark-Pomeron coupling picture.
In this work we extend the framework of Ref. \cite{yu-kong-pin} to
apply for $KN$ ($\overline{K}N$) scattering via four elastic channels,
\begin{eqnarray}
&&K^+p\to K^+p\,,\label{k+p}\\
&&K^-p\to K^-p\,,\label{k-p}\\
&&K^+n\to K^+n\,,\label{k+n}\\
&&K^-n\to K^-n\,,\label{k-n}
\end{eqnarray}
and two charge exchange processes
\begin{eqnarray}
&&K^+n\to K^0p\,,\label{k+ncex}\\
&&K^-p\to \overline{K}^0n\,,\label{k-pcex}
\end{eqnarray}
respectively.

The paper is organized as follows. In Sec. II we begin with
construction of the Reggeized meson exchange for
elastic and charge exchange $KN\,(\overline{K}N)$ scatterings,
excluding $\rho+\omega$
exchanges as discussed in the introduction.
The $t$-channel meson exchanges are applied to elastic
$KN\,(\overline{K}N)$ scattering
to describe the reaction at the Regge realm.
Numerical consequences are compared to
experimental data on total and differential
cross sections at high momentum region. Section III
is devoted to an analysis of the $KN\,(\overline{K}N)$
charge exchange reaction to reproduce experimental data
of total and differential cross sections and beam polarization
asymmetry.
Summary and discussion follow in Sec. IV to evaluate physical
meaning of our findings in the present approach.
In the Appendix we present a numerical evidence for
elastic $KN\,(\overline{K}N)$ cross section with and without
$\rho+\omega$ exchanges to support the present approach.

\section{Elastic $KN\,(\overline{K}N)$ scattering in the Reggeized model}

In the kaon elastic scattering process on nucleon target,
\begin{eqnarray}
K^\pm(k)+ N(p)\to K^\pm(q)+ N(p'),
\end{eqnarray}
the incoming and outgoing kaon momenta are denoted by $k$ and $q$, and the
initial and final nucleon momenta by $p$ and $p'$, respectively.
Then, conservation of four-momentum requires $k+p=q+p'$, and
$s=(k+p)^2$, $t=(q-k)^2$, and $u=(p'-k)^2$ are invariant
Mandelstam variables corresponding to each channel.

Within the present framework where the relativistic Born amplitudes are
employed to be reggeized, the exchanges of $\rho$ and $\omega$
are discarded, as discussed above.
Instead, we consider vector meson $\phi(1020)$ of $J^{PC}=1^{--}$
to assign the role generally expected from the vector meson
exchanges in the $K^\pm N\to K^\pm N$ process.
Of the parity and $C$-parity all
even, the scalar mesons $f_0(980)$, $a_0(980)$ and tensor mesons
$f_2(1270)$, $a_2(1320)$ decaying to $ K\overline{K}$ are
included. For the Pomeron exchange we utilize the amplitude in
Ref. \cite{yu-kong-pin} to describe the reaction cross sections in
the momentum region $P_{\rm Lab}=100 - 200$, GeV/$c$.
Therefore, we write the elastic scattering amplitudes as,
\begin{eqnarray}
&&{\cal M}(K^\pm
p)={f_0}+{a_0}\mp{\phi}+{f_2}+{a_2}+{\mathbb{P}},\label{elasticp}\\
&&{\cal M}(K^\pm
n)={f_0}-{a_0}\mp{\phi}+{f_2}-{a_2}+{\mathbb{P}},\label{elasticn}
\end{eqnarray}
where the $\phi$ meson of $C$-parity odd changes sign between $K^+$ and $K^-$
projectiles and the isovector mesons $a_0$ and $a_2$ change
signs between proton and neutron targets by isospin symmetry.

From the isospin relations between the above amplitudes,
\begin{eqnarray}\label{k+ncex}
&&{\cal M}(K^+n\to K^0p)\nonumber\\&&\hspace{1cm}={\cal M}(K^+p\to
K^+p)-{\cal M}(K^+n\to K^+n),\ \ \ \ \ \\
\label{k-pcex} &&{\cal M}(K^-p\to
\overline{K}^0n)\nonumber\\&&\hspace{1cm} ={\cal M}(K^-p\to
K^-p)-{\cal M}(K^-n\to K^-n),\ \ \ \ \ \
\end{eqnarray}
the charge exchange amplitudes are given by
\begin{eqnarray}
{\cal M}(K^+ n\to K^0p)={\cal M}(K^-p\to \overline{K}^0
n)=2\left(a_0+a_2\right).\hspace{0.5cm}\label{cexpn}
\end{eqnarray}

Given the Reggeized amplitudes relevant to scalar, vector, and
tensor meson exchanges in Ref. \cite{yu-kong-pin}, we now discuss
the determination of coupling constants of the meson exchange in
the $t$-channel.\
\\

$\bullet\ $ Scalar meson exchange

Scalar mesons are expected to give contributions in the
low momentum region.
As to scalar meson-nucleon coupling constants, we appreciate that
$f_0NN$ and $a_0NN$ are still hypothetical yet.
From the $q\bar{q}$ structure of the scalar meson
the QCD-inspired model such as QCD sum rules  predicts that
$g_{f_0NN}=0$ and $g_{a_0NN}=12$, whereas these are
$g_{f_0NN}=10.3$ and $g_{a_0NN}=-8.5$ in the case of the four
quark state $q^2\bar{q}^2$ structure \cite{erkol}.
On the other hand, in the vector meson photoproduction, $\gamma p\to \phi p$,
the scalar meson nonet is considered
with the mixing angle $\theta_s$ between the singlet
and octet members to write the $SU_f(3)$ relations among $a_0$, $f_0$
and $\sigma$ mesons as \cite{titov11}
\begin{eqnarray}
&&g_{a_0NN}={F+D\over3F-D}{1\over\cos\theta_s}g_{\sigma NN},\\
&&g_{f_0NN}=-\tan\theta_s g_{\sigma NN}.
\end{eqnarray}
Given the mixing angle $\theta_s=-3.21^\circ$, and
$F/D=0.575\pm0.016$ we obtain $g_{a_0NN}=31.77\pm1.66$ and
$g_{f_0NN}=0.82$ from $g_{\sigma NN}=14.6$ \cite{bgyu-phi}.
Therefore, within the uncertainty in the choice of $g_{\sigma NN}$
the coupling constant $g_{a_0NN}$ is in the range $8.5 - 32$,
regardless its sign. We choose $g_{a_0NN}=15.5$ which is better to
describe $KN\ (\overline{K}N)$ charge exchange reactions, and
$g_{f_0NN}\approx0$ which is consistent with QCD sum rule and
vector meson photoproduction as well.

The reggeized amplitude for the scalar meson ($S$) exchange employs
the derivative coupling of the scalar meson to $K\overline{K}$ with
the coupling vertex \cite{yu-kong-pin},
\begin{eqnarray}
\Gamma_{SKK}(q,k)={g_{SKK}\over m_K}\,q\cdot k\ .\label{eq11-v}
\end{eqnarray}
As the full width of scalar meson $a_0$ is in a broader range
$\Gamma(a_0)=50 - 100$ MeV, and no precise measurement of the
partial decay width $\Gamma(a_0\to K\overline{K})$ is available
yet, we have to estimate the partial width from PDG;
$\Gamma(a_0\to K\overline{K})=6.4 - 12.8$ MeV from the ratio
$\Gamma(a_0\to K\overline{K})/\Gamma(a_0\to\eta\pi)=0.183\pm0.024$ with
$\Gamma(a_0\to\eta\pi)=35 -  70$ MeV, which is obtained by
$\Gamma(a_0\to\eta\pi)\Gamma(a_0\to\gamma\gamma)/\Gamma(a_0)=0.21^{+0.08}_{-0.04}$
keV and $\Gamma(a_0\to\gamma\gamma)=0.3\pm0.1$ keV.
By isospin invariance of two decay channels $K^+K^-$ and
$K^0\overline{K}^0$ we further consider the factor of $1/2$ for
the above width to get $\Gamma(a_0\to K^+K^-)=3.2 - 6.4$ MeV.
However, because the scalar meson mass is proximity to
the threshold of $K\overline{K}$ decay, the estimate of coupling
constant is highly
sensitive to what mass is chosen for the
$K\overline{K}$ threshold. From the decay width for the derivative
coupling $SKK$ vertex,
\begin{eqnarray}\label{decay}
\Gamma(S\to K^+K^-)={g_{SKK}^2\over8\pi}{k\left(E_K^2+k^2
\right)^2\over
m_S^2m_K^2},
\end{eqnarray}
where $k$ is three-momentum of the decaying meson in the c.m. frame
and $E_K$ is its energy,
we determine $g_{a_0KK}=\pm 5.33$ with the threshold mass
$m_{a_0}=988$ MeV and $\Gamma(a_0\to K^+K^-)=5.06$ MeV
taken, though we use $m_{a_0}=980$ MeV in the analysis of the
reaction process.
\\

$\bullet\ $ Vector meson exchange

The decay width $\phi\to K^+K^-$ is given by
\begin{eqnarray}
\Gamma(\phi\to K^+K^-)={g^2_{\phi KK}k^3\over6\pi m_\phi^2}\ ,
\end{eqnarray}
and the decay width $\Gamma(\phi\to K^+K^-)=2.1$ MeV taken from
PDG yields $g_{\phi KK}=\pm4.46$. In $\phi$ vector meson
photoproduction \cite{ysoh,titov11} the tensor coupling constant
$g^t_{\phi NN}\approx 0$ could obtain a consensus.
However, the vector coupling constant $g^v_{\phi NN}$
is controversial; the value $g^v_{\phi NN}=-0.24$ was used in Ref.
\cite{titov11} which is consistent with $g_{\phi NN}=-0.25$
by the strict OZI rule \cite{ugm}.
Meanwhile, the analysis of NN and YN scattering yields $g_{\phi NN}=\pm1.12$
\cite{nagels17}, and $\pm3.47$ \cite{nagels20},
confirming evidence for OZI evading process at the $\phi NN$ vertex.
From the universality of $\phi$ vector meson dominance,
we here take $g^v_{\phi NN}=-3.0$,
as a trial, which is close to $g_{\phi KK}$
with $g^t_{\phi NN}=0$ for the $\phi NN$ coupling vertex. We note
that whatever values it could take among those suggested in
literature, its contribution should be insignificant in comparison
to $f_2$ and $a_2$, because of the lower lying trajectory as shown
in Table \ref{tb1}.
\\

$\bullet\ $ Tensor meson exchange

The exchange of tensor meson includes isoscalar $f_2$ meson with the
decay width $\Gamma(f_2\to K\overline{K})=4.29$ MeV which is estimated from
the full width 186.7 MeV with the fraction $4.6\%$.
In the isovector channel the $a_2$ meson is considered with the decay
width $\Gamma(a_2\to K\overline{K})=2.69$ MeV from the full width 109.8 MeV
and the fraction $4.9\%$. The factor of 1/2 is taken into account
in both decay widths by the same reason as in the case
of scalar meson.

By using the decay width for the tensor meson $(T)$ coupling to
$K\overline{K}$,
\begin{eqnarray}\label{decay-tensor}
\Gamma(T\to K^+K^-)={4g^2_{TKK}\over 15\pi} {k^5\over
m^4_{T}},
\end{eqnarray}
we estimate the coupling constants as $g_{f_2KK}=\pm3.53$ and
$g_{a_2KK}=\pm2.45$ from the respective decay widths given above.

To keep consistency with the coupling constants of $f_2NN$ in the
$\pi N$ scattering \cite{yu-kong-pin} we resume
$g^{(1)}_{f_2NN}=6.45$ and $g^{(2)}_{f_2NN}=0$, the latter of
which is rather stringent in order to agree with $\pi N$
polarization observables. The determination of $a_2NN$ coupling is
discussed in photoproduction of charged kaon \cite{bgyu-kaon},
where the SU(3)$_f$ symmetry dictates $g^{(1)}_{a_2NN}=1.4$ and
$g^{(2)}_{a_2NN}=0$ to agree with experimental data.  We keep
these values in the present calculation.

A canonical form of the trajectory
$\alpha_\varphi(t)=\alpha_\varphi'(t-m^2_\varphi)+J$ is considered for
the $\varphi$ Regge-pole of spin-$J$. The slopes of $f_2$ and $a_2$
are taken to be the same with those of exchange degenerate pair
$\omega$ and $\rho$, respectively \cite{yu-kong-pin}.
Nevertheless, the phases of $f_2$ and $a_2$ Regge poles are taken
to be exchange nondegenerate for the elastic process, because of the absence of
$\omega$ and $\rho$ from the present calculation.
The slope of $a_0$ trajectory is assumed to be the same with that of scalar meson
$\sigma$ \cite{yu-kong-pin} as a member of the scalar meson nonet.
The trajectory of $\phi$ is taken from Ref. \cite{collins}.
A summary of physical constants is listed in Table~\ref{tb1} including
the coupling constants and trajectories with the corresponding phase factors.
\\

\begin{table}[t]
\caption{Physical constants and Regge trajectories with the
corresponding phase factors for $K^\pm N\to K^\pm N$. The symbol
$\varphi$ stands for $a_0$, $\phi$, $f_2$ and $a_2$. The meson-baryon
coupling constants for vector meson and tensor meson are
denoted by $g^v_{VNN}(g^t_{VNN})$, and $g^{(1)}_{TNN}(g^{(2)}_{TNN})$,
respectively.}
    \begin{tabular}{c|c|c|c|c}\hline
        Meson & Trajectory($\alpha_\varphi$) & Phase factor & $g_{\varphi KK}$& $g_{\varphi NN}$  \\
        \hline\hline
        $a_0$ & $0.7(t-m_{a_0}^2)$ & $(1+e^{-i\pi \alpha_{a_0}})/2$  &$5.33$  & $15.5$\\%
        $\phi$ &  $0.9t+0.1$ &$(-1+e^{-i\pi \alpha_{\phi}})/2$  & $4.46$ & $-3.0$ ($0$) \\%
        $f_2$ &  $0.9t+0.53$ & $(1+e^{-i\pi \alpha_{f_2}})/2$      & 3.53 & $6.45$ ($0$) \\%
        $a_2$ &  $0.9t+0.43$ & $(1+e^{-i\pi \alpha_{a_2}})/2$      &$-2.45$& 1.4 (0) \\
        \hline
    \end{tabular}\label{tb1}
\end{table}

$\bullet\ $ Pomeron  exchange

In Ref. \cite{yu-kong-pin} for $\pi^\pm p$ elastic scatterings
we constructed the Pomeron exchange arising
from the quark-Pomeron coupling picture.

With the Pomeron trajectory,
\begin{eqnarray}\label{pome-traj}
\alpha_\mathbb{P}(t)=0.12\,t+1.06\,,
\end{eqnarray}
and the quark-pion coupling strength $f_{\pi qq}$ from the
Goldberger-Treiman relation at the quark level, the cross sections
for $\pi^\pm N$ elastic scattering up to $P_{\rm Lab}=250$ ${\rm
GeV}/c$ were reproduced by a single exchange of Pomeron. It is
straightforward to apply the formulation of the Pomeron in Ref.
\cite{yu-kong-pin} to $K^\pm p$ elastic scattering\footnote{There
is the Pomeron coupling to strange quark in the $KN$ reaction in
addition to $u(d)$ quark in the quark loop integral
\cite{yu-kong-pin}. Nevertheless, such a difference between the
two couplings by mass difference is neglected for simplicity, as
the quark masses between strangeness and up(down) quarks are not
widely different in the constituent quarks we adopted here.} with
minor changes, e.g., the quark-kaon coupling constant $f_{Kqq}$,
the quark-Pomeron coupling strength $\beta_s$ with the strange
quark mass $m_q$, instead of those for pion and $d$ quark.

In numerical calculations we take $\beta_u=2.07$ and $\beta_s=1.6$
GeV$^{-1}$, as before \cite{bgyu-phi}, and $m_q=500$ MeV in favor
of the strange quark involved.
Let us now determine the coupling strength $f_{Kqq}$. Unlike the
case of $f_{\pi qq}$, however, the Goldberger-Treiman relation from
the SU(3) symmetry is not likely to give the reliable answer by
the large symmetry breaking. From the phenomenological point of
view the ratio of elastic cross sections for $K^+p$ and $\pi^+p$
at high momenta could be a hint to a determination of $f_{Kqq}$
\cite{godizov}.
At $P_{\rm Lab}=250$ ${\rm GeV}/c$ where there exists only the
Pomeron exchange and all others are assumed to be minimal, the
ratio of cross sections from world data gives
\begin{eqnarray}
{\sigma_{el}(K^+p)\over\sigma_{el}(\pi^+p)}\approx 0.836,
\end{eqnarray}
and, hence,
\begin{eqnarray}\label{ratio}
{|{\cal M}_{el}({K^+p})|\over|{\cal M}_{el}({\pi^+p})|} \approx
{f^2_{Kqq}m_K^2\beta_s\over f^2_{\pi qq}m_\pi^2\beta_d}\approx
\sqrt{0.836},
\end{eqnarray}
which yields $f_{Kqq}=0.82$ by taking $f_{\pi qq}=2.65$
\cite{yu-kong-pin}.
This means that, in order to obtain a better agreement with the
high momentum $K^\pm p$ data, we may well treat the quark-meson
coupling constant $f_{Kqq}$ rather as a parameter around the value
above in the fitting procedure to cross section data.

\subsection{$K^\pm p\to K^\pm p$}

\begin{figure}[]
\centering \epsfig{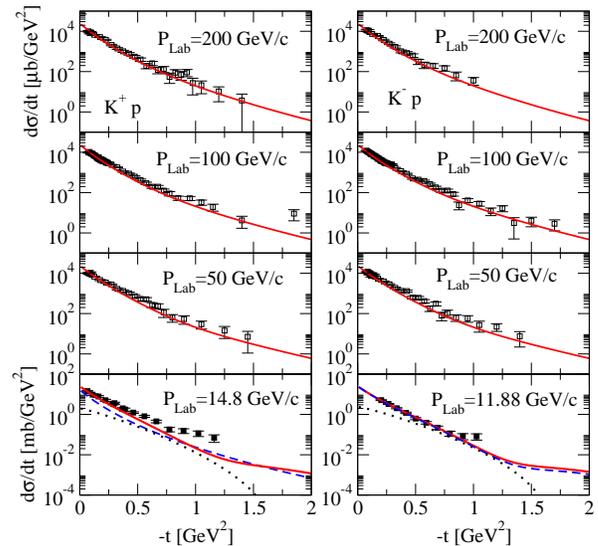}%
\hspace{0.85cm}%
\caption{Differential cross sections for elastic
$K^+p$ (left column) and $K^-p$ (right column). Cross sections at
$P_{\rm Lab}=100$ and 200 ${\rm GeV}/c$ are featured by Pomeron
exchange.
Data at $P_{\rm Lab}=50$, 100, and 200 ${\rm GeV}/c$ for both
reactions are taken from Ref. \cite{akerlof}. Data at $P_{\rm
Lab}=14.8$ ${\rm GeV}/c$ for $K^+p$ and 11.88 ${\rm GeV}/c$ for
$K^-p$ are from Ref. \cite{foley-el}. \\ } \label{fig1}
\end{figure}
\begin{figure}[]
\centering \epsfig{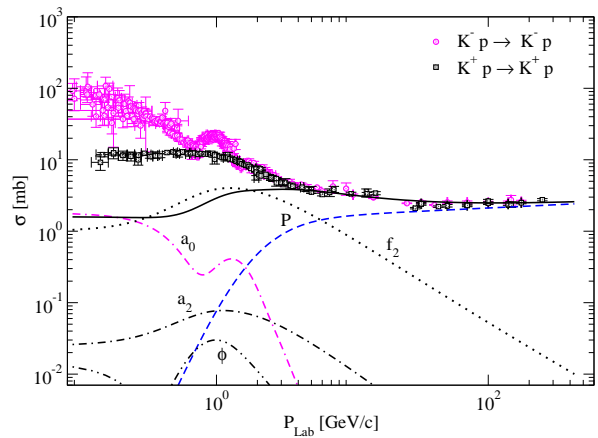}%
\caption{Total elastic cross sections for $K^+p\to K^+p$ and
$K^-p\to K^-p$. The solid curve for the total cross section
results from the full calculation of $K^+p\to K^+p$. The
difference between cross sections of $K^+p$ and $K^-p$ is
negligible by the $\phi$ contribution of the order of $10^{-2}$.
Data are taken from PDG. } \label{fig2}
\end{figure}

Figure \ref{fig1} shows the differential cross sections selected
in the same momentum range to compare $K^+p$ with $K^-p$
elastic scattering. Since the data at $P_{\rm Lab}=100$ and 200
${\rm GeV}/c$ are governed by a pure Pomeron exchange we exploit
them to determine the quark-Pomeron coupling strength, and obtain
$f_{Kqq}=0.988$ with the trajectory in Eq. (\ref{pome-traj}),
while the cutoff parameter $\mu=2.5$ GeV and $n=1$ are fixed for the kaon
form factor \cite{yu-kong-pin},
\begin{eqnarray}\label{kaonff}
F_K(t,W)=\left(1-t/\Lambda^2\right)^{-n}
\end{eqnarray}
and $\Lambda(W)={k\over\mu}(W-W_{th})$.
In a good agreement with all the data selected, we regard
$f_{Kqq}=0.988$ to be reasonable because it is close to 0.82 from
the ratio in Eq. (\ref{ratio}). The dominance of the Pomeron
exchange followed by the tensor meson $f_2$ is shown in the lowest
panels. Contributions of $f_2$ (dotted) and Pomeron (dashed) are
shown at $P_{\rm Lab}=14.8$ for $K^+p$ and 11.88 ${\rm GeV}/c$ for
$K^-p$ reactions, respectively.

In Fig. \ref{fig2} we present total cross sections for $K^+p$ and
$K^-p$ elastic processes where an agreement with data  at high
momenta $P_{\rm Lab}\approx3$ GeV/$c$ is obtained by the exchanges
of $f_2$ and Pomeron. Without a fitting procedure for hadron
coupling constants we describe elastic scattering over the region
$P_{\rm Lab}\approx 3$ ${\rm GeV}/c$ up to 250 ${\rm GeV}/c$.
As discussed above, the distinction between the two reactions
disappears at high momenta, because of the small contribution of
$\phi$. It is worth observing the similarity between $\pi^\pm N$
\cite{yu-kong-pin} and $K^\pm N$ total elastic cross sections
where the roles of $\sigma$, $\rho$ and $\omega$ in the former
reactions are replaced by those of $a_0$, $a_2$ and $\phi$ of the
same order of magnitude, respectively. Also, the $K^\pm p$ elastic
scatterings are dominated by the tensor meson $f_2$ at
intermediate and the Pomeron exchange at high momenta to exhibit
the isoscalar nature of the reactions.

Below $P_{\rm Lab}\approx3$ ${\rm GeV}/c$ the large discrepancy
between the $t$-channel Regge predictions and data could
presumably be resolved by considering nuclear interaction
for $K^+p$ \cite{martin}, and the $\overline{K}N$
coupled states for the $K^-p$ elastic scatterings, respectively \cite{aoki}.

\subsection{$K^\pm n\to K^\pm n$}

\begin{figure}[]
\centering \epsfig{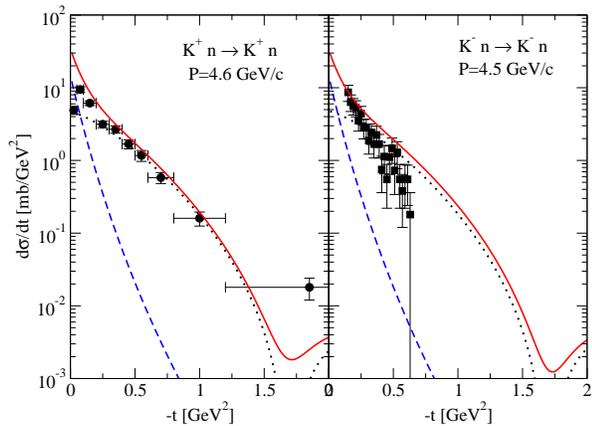}%
\caption{Differential cross sections for $K^+n\to K^+n$ at $P_{\rm
Lab}=4.6$ ${\rm GeV}/c$ and $K^-n\to K^-n$ at $P_{\rm Lab}=4.5$
${\rm GeV}/c$, respectively. Notations for the curves are the
same as in Fig. \ref{fig2}. Data are taken from Refs.
\cite{dehm} and \cite{declercq}, respectively.} \label{fig3}
\end{figure}

Experimental data on neutron targets are extracted from kaon
scattering off deuteron, in which case proton is assumed to be a
spectator. Data at high momenta are rare and contain some
uncertainties due to the procedures such as impulse approximation,
Glauber screening, and Fermi motion taken usually in the analysis
of deuteron data. A few data points on the total cross section for
$K^+n$ elastic reaction are found at $P_{\rm Lab}=2.97$ GeV/$c$
\cite{buchner} and 4.6 GeV/$c$ \cite{dehm}, respectively. For the
$K^-n$ reaction the total cross section data are reported at
$P_{\rm Lab}=2.2$ GeV/$c$ \cite{declais} and 4.5 GeV/$c$
\cite{declercq}, respectively. Therefore, no data are enough to
analyze high momentum behavior of the reactions. Similar to $K^\pm
p$ elastic processes, however, we expect that the difference
between $K^+n$ and $K^-n$ elastic reactions is negligible due to
the mentioned role of vector meson $\phi$.
Furthermore, as the $a_0+a_2$ exchanges in the elastic $K^\pm n$
reaction play the role
opposite to the $K^\pm p$ from Eqs. (\ref{elasticp}) and (\ref{elasticn}),
the difference between them could be less apparent,
and the total elastic
cross sections for $K^\pm n$ at high momentum are to be the same as
those of $K^\pm p$ with the same role of the isoscalar exchanges
$f_2$ and Pomeron. We present differential cross sections for
$K^+n$ at $P_{\rm Lab}=4.6$ ${\rm GeV}/c$ and $K^-n$ at $4.5$
${\rm GeV}/c$ in Fig. \ref{fig3}. We use the kaon form factor with
$\mu=2.5$ GeV and $n=3$ for the Pomeron exchange
in Eq. (\ref{kaonff}) for $K^\pm n$
elastic reactions. An overall agreement with differential cross
sections is predicted, though the discrepancy with $K^+n$ data in
the large $-t$ is shown due to the dominance of $f_2$ exchange
with the exchange nondegenerate phase over the Pomeron
contribution in these intermediate momenta, $4.5$ and 4.6
GeV/$c$.

\section{Charge exchange scattering}

\subsection{$K^+n\to K^0 p$ and $K^-p\to \overline{K}^0n$}

\begin{figure}[]
\centering \epsfig{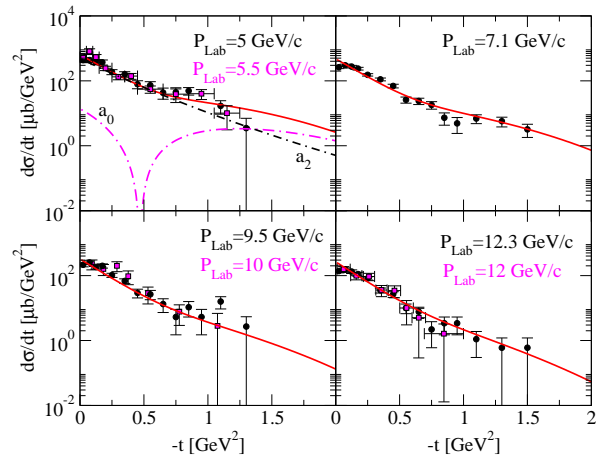}%
\caption{Differential cross sections for $K^- p\to
\overline{K}^0n$ and $K^+n\to K^0p$. Solid curves for cross
sections are calculated at $P_{\rm Lab}=5.5$, 7.1, 10, and 12
GeV/$c$ by using the $K^+p\to K^0n$ amplitude, which shares with
the $K^-p\to \overline{K}^0n$ amplitude in common. The respective
contributions of $a_0(980)$ and $a_2(1320)$ are shown in the upper
left panel. Data of $K^-p\to \overline{K}^0n$ (black circles) are
taken from Ref. \cite{astbury}. Data of $K^+n\to K^0p$ (magenta
squares) at $P_{\rm Lab}=5.5$, 10, and 12 ${\rm GeV}/c$ are taken
from Refs. \cite{cline}\cite{haguenauer}\cite{firestone}. \\ \\ }
\label{fig4}
\end{figure}
\begin{figure}[]
\centering \epsfig{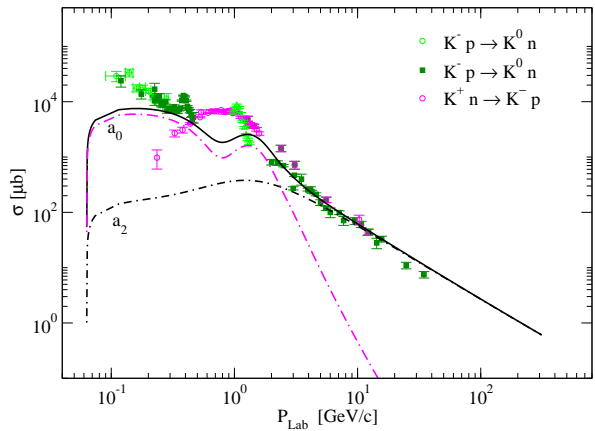}%
\caption{Total charge exchange cross sections for $K^- p\to
\overline{K}^0n$ and $K^+n\to K^0p$. The solid cross section is
presented for the $K^+n\to K^0p$ process. The contribution of the
$a_2$ exchange agrees with data at high momenta. Data of $K^+n\to
K^0p$ are taken from Refs. \cite{haguenauer,damerell75}. Data of
$K^- p\to \overline{K}^0n$ are from Refs.
\cite{astbury,mast,abrams,dehm,foley-tcs,conforto}. \\ }
\label{fig5}
\end{figure}
\begin{figure}[]
\centering \epsfig{file=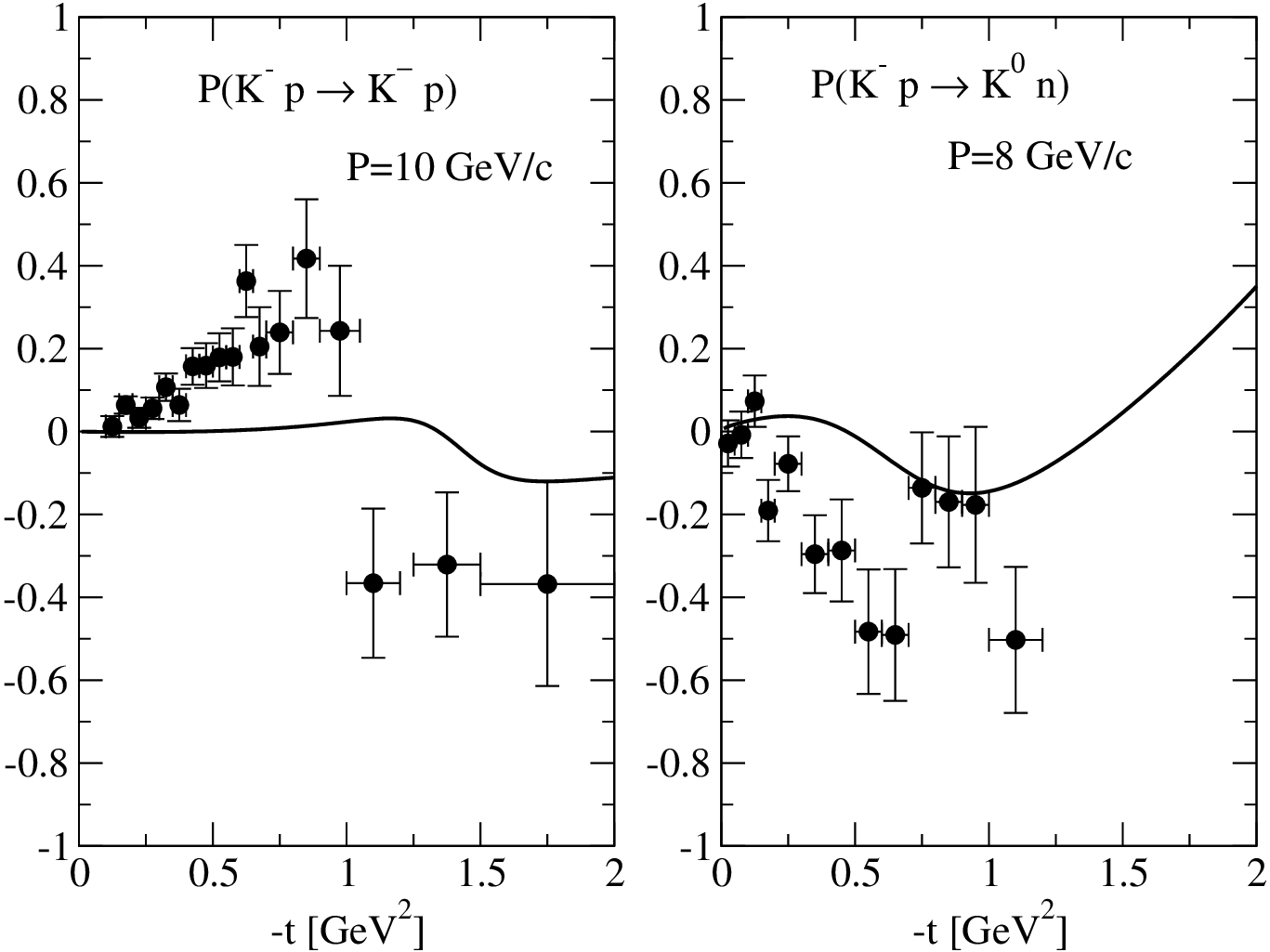, width=0.9\hsize}%
\caption{Polarization $P$ of $K^-p\to K^-p$ at $P_{\rm
Lab}=10$ GeV/$c$ and $K^-p\to\overline{K}^0n$ at $P_{\rm
Lab}=8$ GeV/$c$. Data are taken from Refs. \cite{borghini,beusch}.}
\label{fig6}
\end{figure}

Charge exchange $K^+n$ and $K^-p$ are good places to examine the
validity of simple $a_0+a_2$ exchanges. A collection of data
on these reactions exhibits the same dependence of both cross sections
upon energy and angles at high momenta, as shown in Figs.
\ref{fig4} and \ref{fig5}. The differential cross
sections for both reactions in the same momentum range
are presented in Fig. \ref{fig4}. We obtain a fair agreement with
data by using the coupling constants and the exchange nondegenerate
phase for $a_0$ chosen in Table \ref{tb1} for the combined analysis.
However, in these reactions, the complex
phase, EXP$[-i\pi\alpha_{a_2}(t)]$ for the $a_2$ exchange
is favored in order to agree with cross section data rather than
the exchange nondegenerate phase.
In Fig. \ref{fig5} such an expected behavior from the $a_2$ exchange
with the complex phase is clear in the total cross section, where the
cross section over $P_{\rm Lab}\approx4$ GeV/$c$
should be reproduced only by the $a_2$ exchange.
Meanwhile, the $a_0$ exchange is found to play a leading
role in the low momentum region.
Throughout the successful description of reaction
cross sections as presented in Figs. \ref{fig4} and \ref{fig5},
it could be concluded that the isospin symmetry is valid between
all six channels as stated in Eqs. (\ref{k+ncex}) and (\ref{k-pcex}).

Finally, we discuss polarization $(P)$ of $KN\,(\overline{K}N)$ scattering.
It is observed via the interference between the spin-nonflip and
spin-flip amplitudes of exchanged mesons \cite{yu-kong-pin}.
Therefore, we simply figure out vanishing of polarization at high momenta,
because there contributes only Pomeron exchange.
Within the present framework the predictions for polarization data
at intermediate momenta are poor, as shown by the case of $K^- p$ elastic
reaction in Fig. \ref{fig6}.
Nevertheless, due to the interference between $a_0$ with the exchange
nondegenerate phase and $a_2$ with the complex phase the polarization of
$K^-p\to\overline{K}^0n$ at $P_{\rm Lab}=8$ GeV/$c^2$
is reproduced to some degree. The polarizations in both
reactions follow the behavior of data along with $t$-dependence in Fig. \ref{fig6}.
Inclusion of the cut is likely to reinforce the strength of the polarization
to agree with data, as demonstrated in Ref. \cite{yu-kong-pin}.

\section{summary and discussion}

We have performed a combined analysis of $KN$ and $\overline{K}N$
scatterings with a set of coupling constants common in four
elastic and two charge exchange processes in the Regge realm. The
scattering amplitude is obtained by reggeizing the relativistic
Born amplitude for the $t$-channel meson exchange with the decay
width to $K\overline{K}$ allowed kinematically and listed in the PDG.
Thus, the exchanges of
$\rho$ and $\omega$ are excluded from the present framework and
the present approach needs none of the hadron form factors for such
off-shell meson exchanges.
This is the most distinctive feature from previous calculations at
high momenta, though contradicting to existing model
descriptions.

Within the present approach
the isoscalar $f_2$ and Pomeron exchanges are found to be dominant
in four elastic channels, while the charge exchange processes
exhibit the isovector nature via the simple $a_0$ and $a_2$ exchanges.
The exchange of the soft Pomeron which is constructed on the basis of
the quark-Pomeron coupling picture reproduces the diffraction feature
of elastic cross sections to a good degree.
Nevertheless, in order for the present
analysis to be valid for the threshold region,
inclusion of partial wave contributions
is necessary to describe the reaction mechanisms induced by either the
propagation of exotic channels in $K^+N$ reactions, or the
meson-baryon couplings in the
$\overline{K}N$ channels. This should be a subject of future
study to make complete our understanding of $KN\,(\overline{K}N)$
scattering based on the $t$-channel exchange discussed
here as a background contribution.
Work on this direction is ongoing and the results will appear elsewhere.

       \section*{Acknowledgments}
This work was supported by the National Research Foundation of
Korea Grant No. NRF-2017R1A2B4010117.

\appendix

\section{Conventional Approach versus the Reggeized model}

It has been a long standing idea that the $\rho$ and $\omega$
vector meson exchanges are included to account for the difference
between $K^+p$ and $K^-p$ total cross sections at high momenta.
\cite{donnachie}. At this point, the elastic cross section should
not be confused with the total (reaction) cross section. The
former reaction cross sections are observed to coincide to each
other at high momenta, which is understandable only by the
isoscalar Pomeron exchange.

\begin{figure}[b]
\centering \epsfig{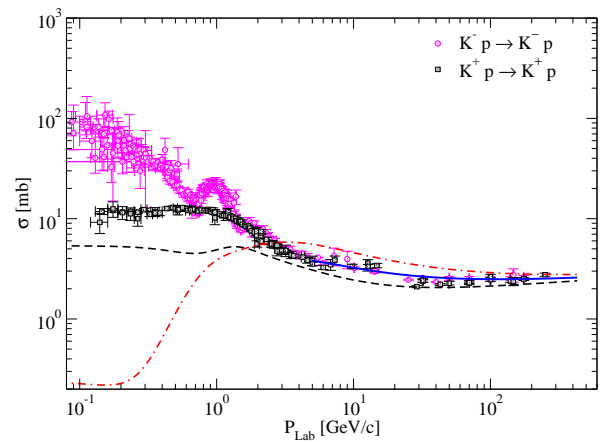}%
\caption{Effect of $\rho+\omega$ exchanges on the elastic $K^\pm
p$ cross sections. The discrepancy between $K^-p$ (upper
dash-dotted) and $K^+p$ (lower dashed) curves at high momenta
cannot be reduced to a common datasets for $K\pm p$ in the
presence of $\rho+\omega$ exchanges. The solid curve from Fig.
\ref{fig2} is presented for comparison. } \label{fig0}
\end{figure}

For illustration purpose, we make a simulation of $\rho+\omega$
exchanges, despite the decay mode neither $\rho\to
K\overline{K}$ nor $\omega\to K\overline{K}$ is allowed. These vector
mesons could be considered in the elastic $K^\pm N$ scattering by
substituting
$f_2\to (f_2\mp\omega)$ and $a_2\to (a_2\mp\rho)$ in Eqs.
(\ref{elasticp}) and (\ref{elasticn}) \cite{phillips} with the
exchange degenerate phase $(a_2\mp\rho)=1$, or
$e^{-i\pi\alpha(t)}$ for the minus, or plus sign.
The trajectories $\alpha_\rho(t)=0.9t+0.46$ and
$\alpha_\omega(t)=0.9t+0.44$ are used. To be consistent with
other meson exchanges we
avoid employing cutoff form factors in the meson baryon coupling
vertices. However, given the coupling constants $g_{\rho
KK}=g^v_{\rho NN}$, and $g_{\omega KK}=g^v_{\omega NN}$, which we
usually take 2.6 and 15.6, respectively, from vector meson
dominance \cite{butt,bszou} the inclusion of $\rho+\omega$
exchanges lead to a complete failure in
reproducing elastic data. Figure \ref{fig0} shows inconsistency of
model predictions for the $K^\pm p$ elastic cross sections, even
if we use such unnatural values $g_{\rho KK}=0.1$ and $g_{\omega
KK}=-0.1$, while the vector-meson nucleon coupling constants
remain unchanged.

As demonstrated by the solid curve for
reference, the coincidence of $K^\pm p$ cross sections at high
momenta can be achieved only from the absence of $\rho+\omega$
exchanges either or they should be negligible at least within the
present approach. To investigate the possibility of spin-1
vector meson further, we test the contribution of $\rho'(1450)$
exchange with the
trajectory $\alpha_{\rho'}=t-1.23$. As the coupling constants are
still evasive, we deduced $G^v_{\rho'}\approx 10$ and
$G^t_{\rho'}\approx-20$ from those values, 40 and $-75$, which are
corresponding to the $\pi N$ case \cite{yu-kong-pin}. Assuming the
universality, $g_{\rho'\pi\pi}=g^v_{\rho'NN}\approx6.32$, and we
take half the value of it for the present case with
$g_{\rho'KK}=g^v_{\rho'NN}$ by the same reason. From the ratio of
vector to tensor coupling, $\kappa_{\rho'\pi\pi}\approx-1.88$ is
deduced. The result hardly alters the cross section in the
momentum region of our interest. These findings show that they are
inadequate for the present framework which is
based on nearly on-shell Born amplitude for the reggeization
of $t$-channel meson exchange.


\end{document}